\documentclass [print,amsmath,amssymb,11pt]{revtex4}
\usepackage{amsthm}
\usepackage{charter}
\usepackage{graphicx}
\usepackage{dcolumn}
\usepackage{bm}
\usepackage{dsfont}
\usepackage[landscape,papersize={297.1mm,210mm},left=1.9cm,right=1.6cm,top=2.7cm,bottom=2.8cm]{geometry}% 此指令是规定A4纸的页边距
\usepackage[
            pdfstartview=FitH,
            colorlinks, %注释掉此项则交叉引用为彩色边框(将colorlinks和pdfborder 同时注释掉)
            pdfborder=001,   %注释掉此项则交叉引用为彩色边框
            linkcolor=blue,
            anchorcolor=blue,
            citecolor=blue,
            urlcolor=blue
            ]{hyperref}

\newcommand{\T}{\mbox{$\mathrm{tr}$}}
\usepackage{epstopdf}
\newtheorem{theorem}{Theorem}

\begin{document}
\title{ Tighter constraints of multiqubit entanglement in terms of unified entropy}
\author{Ya-Ya Ren$^{1}$}
\email{yysz7900@163.com}
\author{Zhi-Xi Wang$^1$}
\email{wangzhx@cnu.edu.cn}
\author{Shao-Ming Fei$^{1,2}$}
\email{feishm@cnu.edu.cn}

\affiliation{$^1$School of Mathematical Sciences, Capital Normal University, Beijing 100048, China\\
$^2$Max-Planck-Institute for Mathematics in the Sciences, 04103, Leipzig, Germany}

\begin{abstract}
We present classes of monogamy inequalities related to the $\alpha$-th ($\alpha \geq 1$) power of the entanglement measure based on the unified-($q,s$) entropy, and polygamy inequalities related to the $\beta$-th ($0 \leq \beta \leq 1$) power of the unified-($q,s$) entanglement of assistance by using Hamming weight. We show that these monogamy and polygamy inequalities are tighter than the existing ones. Detailed examples are given for illustrating the advantages.

\end{abstract}

\maketitle

\section{Introduction}
As an important physical resource the quantum entanglement is the most intrinsic feature of quantum mechanics and plays a central role in quantum information processing \cite{c0,c1,c2,c3,b1,b2}.
An important property of quantum entanglement is the monogamy \cite{c4}, which may play significant roles in quantum key distribution security \cite{c6}.
For a tripartite quantum state $\rho_{ABC}$, the usual monogamy inequality can be mathematically characterized as $\mathcal{E}(\rho_{A|BC})\geq\mathcal{E}(\rho_{A|B})+\mathcal{E}(\rho_{A|C})$,
where $\rho_{A|B}=\T_C(\rho_{ABC})$ and $\rho_{A|C}=\T_B(\rho_{ABC})$ are reduced density matrices of the quantum state $\rho_{ABC}$, $\mathcal{E}$ stands for an entanglement measure of bipartite states. However, not all entanglement measures satisfy such monogamy inequality for all quantum states. Coffman-Kundu-Wootters (CKW) first proved that the squared concurrence $C^2$ satisfies the monogamy inequality for arbitrary three-qubit states \cite{c14}.
Such three-qubit CKW inequality was then generalized to the case of arbitrary multiqubit systems \cite{c15}. Later on, various monogamy relations have been presented for a variety of bipartite entanglement measures such as entanglement negativity \cite{NEG1,NEG2,NEG3}, Tsallis-$q$ entanglement \cite{c18,b4} and R\'enyi entanglement \cite{KSRenyi}.

The monogamy of entanglement refers to the fact that the restricted sharability of entanglement in a composite system. As the dual relations to monogamy, the polygamy relations can be characterized as $E_a(\rho_{A|BC})\leq E_a(\rho_{A|B})+E_a(\rho_{A|C})$ for tripartite state $\rho_{ABC}$, where $E_a$ is the assisted entanglement \cite{b3}, a concept dual to bipartite entanglement measures. The dual CKW inequality was proposed for three-qubit systems by using the concurrence of assistance \cite{b3}. Later, polygamy inequalities have been further generalized to multiqubit systems under various entropic entanglement measures \cite{c18,c19}.
General polygamy inequalities of entanglement have been established for multipartite quantum systems with arbitrary dimensions \cite{c20,c21}. Recently, various monogamy and polygamy inequalities of entanglement have been derived in terms of non-negative power of multipartite entanglement measures \cite{zhu,JSK7,jin,jin3,jin4}.

In this paper, we study the monogamy and polygamy relations based on the unified-($q,s$) entropy \cite{b27,b28} for multiqubit systems.
By using the Hamming weight of binary vectors we obtain a class of tight monogamy inequalities for multiqubit entanglement in terms of the $\alpha$-th power of unified entanglement for $\alpha  \geq 1$.
We also present a class of tight polygamy inequalities in terms of the $\beta$-th power of the unified entanglement of assistance for $0 \leq \beta \leq 1$.
We show that these monogamy and polygamy relations are tighter than the existing ones. Moreover, for $\alpha<0$, we obtain a upper bound of unified entanglement in terms of the $\alpha$-th power.

\section{Preliminaries}
We first recall the conceptions of unified entropy, unified-$(q, s)$ entanglement and unified-$(q, s)$ entanglement of assistance as well as the monogamy and polygamy inequalities based on unified entanglement for multiqubit systems.

The unified-$(q,s)$ entropy of a quantum state $\rho$ is defined as \cite{b27,b28}
\begin{equation*}
S_{q, s}(\rho):=\frac{1}{(1-q)s}\left[{\left(\T \rho^{q}\right)}^s-1\right],
\end{equation*}		
with $q, s \geq 0$, $q \neq 1$ and $s \neq 0$.
Although unified-$(q, s)$ entropy has singularities at $s=0$ or $q=1$, it
converges to the R\'enyi-$q$ entropy as $s$ tends to 0 \cite{renyi}, and to Tsallis-$q$ entropy \cite{tsallis} when $s$ tends to $1$.
For $s\geq0$, the unified-$(q, s)$ entropy converges to the von Neumann entropy $S(\rho)$ when $q$ tends to 1, $\lim_{q \rightarrow 1}S_{q, s}(\rho)=-\T \rho\log_{2}\rho=:S(\rho)$.

The unified-$(q, s)$ entanglement (UE) $E_{q, s}\left(|\psi \rangle_{AB} \right)$ of a bipartite pure state $|\psi \rangle_{AB}$ is defined as \cite{c29}
\begin{equation*}
E_{q, s}\left(|\psi \rangle_{A|B} \right)=S_{q, s}(\rho_A),
\label{pure}
\end{equation*}
for each $q, s \geq 0$, where $\rho_A= \T_B (|\psi \rangle_{AB} \langle \psi|)$ is the reduced density matrix of system $A$.
The unified-$(q, s)$ entanglement of a bipartite mixed state $\rho_{AB}$ is given by
\begin{equation}
E_{q, s}\left(\rho_{A|B} \right)=\min \sum_i p_i E_{q ,s}(|\psi_i \rangle_{A|B}), \label{n1}
\end{equation}
where the minimum is taken over all possible pure state decompositions of
$\rho_{AB}=\sum_{i}p_i (|\psi_i \rangle_{AB} \langle \psi _i|)$ with $p_i \geq 0$ and $\sum_{i}p_i=1$.

As a dual concept to UE, the unified-$(q, s)$ entanglement of assistance (UEoA) is defined as \cite{c30}
\begin{equation}
E^{a}_{q, s}\left(\rho_{A|B} \right)=\max \sum_i p_i E_{q, s}(|\psi_i \rangle_{A|B}), \label{n2}
\end{equation}
for each $q, s \geq 0$, where the maximum is taken over all possible ensemble of $\rho_{AB}=\sum_{i}p_i (|\psi_i \rangle_{AB} \langle \psi _i|)$ with $p_i \geq 0$ and $\sum_{i}p_i=1$ .

Due to the continuity of UE in Eq. \eqref{n1} with respect to the parameters $q$ and $s$, UE reduces to the one-parameter class of entanglement measures named R\'enyi-$q$ entanglement \cite{KSRenyi} as $s$ tends to $0$,
and to the bipartite entanglement measures called Tsallis-$q$ entanglement (TE) \cite{c18}
as $s$ tends to $1$.
For any nonnegative $s$, UE reduces to the entanglement of formation as $q$ tends to 1,
$\lim_{q\rightarrow1}E_{q,s}\left(\rho_{A|B} \right)=E_{f}\left(\rho_{A|B} \right)$.
Thus, UE is one of the most common classes of bipartite entanglement measures, including the classes of the R\'enyi, Tsallis entanglement and the entanglement of formation as special cases \cite{c29}.

Similarly, due to the continuity of UEoA in Eq. \eqref{n2} with respect to the parameters $q$ and $s$, when $s$ tends to $0$ or $1$, UEoA is reduced to the R\'enyi-$q$ entanglement of assistance \cite{KSRenyi} and the Tsallis-$q$ entanglement of assistance \cite{c18}, respectively. For any nonnegative $s$, UEoA degenerates into the entanglement of assistance \cite{c30} when $q$ tends to $1$, $\lim_{q\rightarrow1}{E^a_{q, s}}\left(\rho_{A|B}
\right)=E^a\left(\rho_{A|B} \right)$.

The UE in Eq. \eqref{n1} satisfies the following monogamy inequality for any multiqubit state $\rho_{A B_0 \cdots B_{N-1}}$ \cite{c29},
\begin{equation}
E_{q, s}\left( \rho_{A|B_0 \cdots B_ {N-1}}\right)\geq \sum_{i=0}^{N-1}E_{q, s}\left(\rho_{A|B_i}\right),\label{n7}
\end{equation}
where $q\geq2$, $0\leq s \leq1$ and $qs\leq3$, $E_{q, s} ( \rho_{A|B_0 \cdots B_{N-1}})$ is the UE of $\rho_{A B_0 \cdots B_N-1}$ under the bipartite partition $A$ and $B_0 \cdots B_{N-1}$, $E_{q, s}\left(\rho_{A| B_i}\right)$ is the UE of the reduced state $\rho_{A B_i}$ for $i=0,\cdots,N-1$.

Later, a class of polygamy inequalities has been proposed for multiqubit systems in terms of UEoA \cite{c30},
\begin{equation}
E^{a}_{q, s}\left( \rho_{A|B_0 \cdots B_ {N-1}}\right)\leq \sum_{i=0}^{N-1}E^{a}_{q, s}\left(\rho_{A|B_i}\right),
\label{n8}
\end{equation}
for $1 \leq q \leq 2$ and $-q^2+4q-3 \leq s \leq1$, where $E^{a}_{q, s} ( \rho_{A B_0 \cdots B_{N-1}})$ is the UEoA of $\rho_{A B_0 \cdots B_N-1}$ with
respect to the bipartite partition $A$ and $B_0 \cdots B_{N-1}$, $E^{a}_{q, s}\left(\rho_{A| B_i}\right)$ is the UEoA of the reduced state $\rho_{A B_i}$, $i=0,\cdots,N-1$.

In \cite{b26}, Kim put forward the concept of Hamming weight. For any nonnegative integer $j$ with it's binary expansion
$j=\sum_{i=0}^{n-1} j_i 2^i$, where $\log_{2}j \leq n$ and $j_i \in \{0, 1\}$, $i=0, \cdots, n-1$,  a unique binary vector associated with $j$ can be defined as
$\overrightarrow{j}=\left(j_0,~j_1,~\cdots ,j_{n-1}\right)$.
From the above definition of the binary vector $\overrightarrow{j}$ of $j$, one defines the Hamming weight $\omega_{H}\left(\overrightarrow{j}\right)$ as the number of $1's$ in the coordinates of $\overrightarrow{j}$ \cite{c0}.
Thus, one has
\begin{equation}\label{n11}
\omega_{H}\left(\overrightarrow{j}\right)\leq j.
\end{equation}
Kim proposed a class of tight constrains on the $\alpha$-th power of multiqubit entanglement based on the Hamming weight \cite{b26}. When $q\geq2$, $0\leq s \leq1$ and $qs\leq3$,
\begin{equation}\label{n12}
[E_{q, s}(\rho_{A|B_0B_1\ldots B_{N-1}})]^\alpha \geq \sum\limits_{j=0}^{N-1} \alpha^{\omega_{H}(\vec{j})}[E_{q, s}(\rho_{A|B_j})]^\alpha
\end{equation}
for $\alpha \geq 1$, and when $1 \leq q \leq 2$ and $-q^2+4q-3 \leq s \leq1$,
\begin{equation}\label{n13}
[E_{q, s}^a(\rho_{A|B_0B_1\ldots B_{N-1}})]^\alpha \leq \sum\limits_{j=0}^{N-1}\alpha^{\omega_{H}(\vec{j})}[E_{q, s}^a(\rho_{A|B_j})]^\alpha
\end{equation}
for $0\leq\alpha \leq 1$.
Inequalities \eqref{n12} and \eqref{n13} can be further written as
\begin{equation*}\label{n14}
[E_{q, s}(\rho_{A|B_0B_1\ldots B_{N-1}})]^\alpha \geq \sum\limits_{j=0}^{N-1} \alpha ^{j}[E_{q, s}(\rho_{A|B_j})]^\alpha
\end{equation*}
for $\alpha \geq 1$, and
\begin{equation*}\label{n15}
[E_{q, s}^a(\rho_{A|B_0B_1\ldots B_{N-1}})]^\alpha \leq \sum\limits_{j=0}^{N-1}\alpha ^{j}[E_{q, s}^a(\rho_{A|B_j})]^\alpha
\end{equation*}
for $0 \leq \alpha \leq 1$
under additional conditions $E_{q, s}(\rho_{A|B_i}) \geq \sum\limits_{j=i+1}^{N-1}E_{q, s}(\rho_{A|B_j})$ for $i=0,\dots,N-2$ and
$E_{q, s}^a(\rho_{A|B_i}) \geq \sum\limits_{j=i+1}^{N-1}E_{q, s}^a(\rho_{A|B_j})$ for $i=0,\dots,N-2$ respectively.

\section{Tighter monogamy constraints of multiqubit entanglement in terms of UE}

In this section, we provide a class of tighter monogamy relations in terms of the $\alpha$-th power of UE for multiqubit systems.

\begin{theorem}
For any multiqubit state $\rho_{AB_0\ldots B_{N-1}}$, $\alpha\geq1$, $q\geq2$, $0\leq s \leq1$ and $qs\leq3$, we have
\begin{equation}\label{n18}
[E_{q, s}(\rho_{A|B_0B_1\ldots B_{N-1}})]^\alpha
\geq\sum\limits_{j=0}^{N-1}\Big(\frac{(1+k^\delta)^\alpha-1}{k^{\delta\alpha}}\Big)^{\omega_H(\vec{j})}[E_{q, s}(\rho_{A|B_j})]^\alpha,
\end{equation}
where $0<k\leq 1$ and $\delta\geq 1$, $\overrightarrow{j}=\left(j_0, \cdots ,j_{n-1}\right)$ is the vector from the binary representation of $j$ and
$\omega_{H}\left(\overrightarrow{j}\right)$ is the Hamming weight of $\overrightarrow{j}$.
\end{theorem}

{\sf [Proof]}
Without loss of generality, we assume that the qubit subsystems $B_0, \ldots, B_{N-1}$ satisfies the following condition,
\begin{equation}\label{n19}
k^{\delta}E_{q, s}(\rho_{A|B_j})\geq E_{q, s}(\rho_{A|B_{j+1}})\geq 0
\end{equation}
for $j=0,1,\ldots,N-2$ and $0<k\leq1$, $\delta\geq1$ by reordering and relabeling.
From the monotonicity of the function $f(x)=x^{\alpha}$ for $\alpha \geq 1$ and the inequality \eqref{n7} we have,
\begin{align}
\left[E_{q, s}\left(\rho_{A|B_0B_1\cdots B_{N-1}}\right)\right]^{\alpha}\geq
[\sum_{j=0}^{N-1}E_{q, s}\left(\rho_{A|B_j}\right)]^{\alpha}.
\label{n9}
\end{align}
Hence, to prove the theorem we need to prove that
\begin{equation}\label{n20}
\left[\sum\limits_{j=0}^{N-1} E_{q,s }(\rho_{A|B_j})\right]^\alpha
\geq\sum\limits_{j=0}^{N-1}\Big(\frac{(1+k^{\delta})^\alpha-1}{k^{\delta\alpha}}\Big)^{\omega_{H}(\vec{j})}
[E_{q, s}(\rho_{A|B_j})]^\alpha.
\end{equation}

We first prove the inequality \eqref{n20} for $N=2^n$ by using mathematical induction on $n$, and then extend the result to arbitrary positive integer $N$.
For $n=1$ and a three-qubit state $\rho_{AB_0 B_1}$, with the two-qubit reduced density matrices $\rho_{AB_0}$ and $\rho_{AB_1}$, we obtain
\begin{equation}\label{n21}
[E_{q, s}(\rho_{A|B_0})+E_{q, s}(\rho_{A|B_1})]^\alpha
=[E_{q, s}(\rho_{A|B_0})]^\alpha \Big(1+\frac{E_{q, s}(\rho_{A|B_1})}{E_{q, s}(\rho_{A|B_0})}\Big)^\alpha.
\end{equation}
Combining \eqref{n19} with the following inequality \cite{LYY},
   \begin{equation*}
	(1+x)^{\alpha}\geq 1+\frac{(1+k^{\delta})^{\alpha}-1}{k^{\delta\alpha}}x^{\alpha}, \label{n16}
   \end{equation*}
where $k$ and $\delta$ are real numbers such that $0<k \leq 1$ and $\delta\geq 1$, $0\leq x \leq k^{\delta}$ and $\alpha \geq 1$, we have
\begin{equation}\label{n22}
\Big(1+\frac{E_{q, s}(\rho_{A|B_1})}{E_{q, s}(\rho_{A|B_0})}\Big)^\alpha \geq
1+\frac{(1+k^{\delta})^\alpha-1}{k^{\delta\alpha}}\Bigg(\frac{E_{q, s}(\rho_{A|B_1})}{E_{q, s}(\rho_{A|B_0})}\Bigg)^\alpha.
\end{equation}
From \eqref{n21} and \eqref{n22}, we get
\begin{equation}
[E_{q, s}(\rho_{A|B_0})+E_{q, s}(\rho_{A|B_1})]^\alpha\geq
[E_{q, s}(\rho_{A|B_0})]^\alpha+\displaystyle\frac{(1+k^\delta)^\alpha-1}{k^{\delta\alpha}}[E_{q, s}(\rho_{A|B_1})]^\alpha,\label{n10}
\end{equation}
which recovers the inequality \eqref{n20} for $n=1$.

Next, we assume that the inequality \eqref{n20} holds for $N=2^{n-1}$ with $n\geq 2$,
and consider the case of $N=2^n$.
For an $(N + 1)$- qubit state $\rho_{AB_0B_1\ldots B_{N-1}}$,
we have $E_{q, s}(\rho_{A|B_{j+2^{n-1}}})\leq k^{\delta2^{n-1}}E_{q, s}(\rho_{A|B_j})$ from \eqref{n19}.
Therefore,
\begin{equation*}
0\leq\frac{\sum\nolimits_{j=2^{n-1}}^{2^n-1}E_{q, s}(\rho_{A|B_j})}{\sum\nolimits_{j=0}^{2^{n-1}-1}
E_{q, s}(\rho_{A|B_j})}\leq k^{\delta2^{n-1}}\leq k^{\delta}
\end{equation*}
and
\begin{equation*}
\Bigg(\sum\nolimits_{j=0}^{N-1}E_{q, s}(\rho_{A|B_j})\Bigg)^\alpha
=\Bigg(\sum\nolimits_{j=0}^{2^{n-1}-1}E_{q, s}(\rho_{A|B_j})\Bigg)^\alpha
\Bigg(1+\frac{\sum_{j=2^{n-1}}^{2^n-1}E_{q, s}(\rho_{A|B_j})}{\sum_{j=0}^{2^{n-1}-1}E_{q, s}
(\rho_{A|B_j})}\Bigg)^\alpha.
\end{equation*}
Thus, we have
\begin{equation} \label{n23}
\Bigg(\sum\nolimits_{j=0}^{N-1}E_{q, s}(\rho_{A|B_j})\Bigg)^\alpha
\geq \Bigg(\sum\nolimits_{j=0}^{2^{n-1}-1}E_{q, s}(\rho_{A|B_j})\Bigg)^\alpha
+\displaystyle\frac{(1+k^{\delta})^\alpha-1}{k^{\delta\alpha}}\Bigg(\sum\nolimits_{j=2^{n-1}}^{2^n-1}E_{q, s}(\rho_{A|B_j})\Bigg)^\alpha.
\end{equation}
From the induction hypothesis, we get
$$
\Bigg(\sum\nolimits_{j=0}^{2^{n-1}-1}E_{q, s}(\rho_{A|B_j})\Bigg)^\alpha \geq
\sum\nolimits_{j=0}^{2^{n-1}-1}\Big(\frac{(1+k^{\delta})^\alpha-1}{k^{\delta\alpha}}\Big)^{\omega_H(\vec{j})}
[E_{q, s}(\rho_{A|B_j})]^\alpha.
$$
Moreover, the last summation in inequality \eqref{n23} is also a summation of $2^{n-1}$ terms starting from $j=2^{n-1}$ to $j=2^n-1$. Therefore, the induction hypothesis also leads to
$$
\Bigg(\sum\nolimits_{j=2^{n-1}}^{2^n-1}E_{q, s}(\rho_{A|B_j})\Bigg)^\alpha \geq
\sum\nolimits_{j=2^{n-1}}^{2^n-1}\Big(\frac{(1+k^{\delta})^\alpha-1}
{k^{\delta\alpha}}\Big)^{\omega_H(\vec{j})-1}[E_{q, s}(\rho_{A|B_j})]^\alpha.
$$
Thus, we have
\begin{equation}
\Bigg(\sum\nolimits_{j=0}^{2^n-1}E_{q, s}(\rho_{A|B_j})\Bigg)^\alpha\geq
\sum\nolimits_{j=0}^{2^n-1}\Big(\frac{(1+k^{\delta})^\alpha-1}{k^{\delta\alpha}}\Big)^{\omega_H(\vec{j})}[E_{q, s}(\rho_{A|B_j})]^\alpha,\label{n24}
\end{equation}
which recovers the inequality \eqref{n20} for $N=2^n$.

Now let us consider an arbitrary positive integer $N$ and an $(N+1)$-qubit state $\rho_{AB_0B_1\ldots B_{N-1}}$, $0<N \leq 2^n$ for some $n$. We consider a $(2^n+1)$-qubit state,
\begin{equation}\label{n25}
\Gamma_{AB_0B_1\ldots B_{2^n-1}}=\rho_{AB_0B_1\ldots B_{N-1}}\otimes \sigma_{B_N\ldots B_{2^n-1}},
\end{equation}
which is a product of $\rho_{AB_0B_1\ldots B_{N-1}}$ and an arbitrary $(2^n-N)$-qubit state $\sigma_{B_N\ldots B_{2^n-1}}$.
Because $\Gamma_{AB_0B_1\ldots B_{2^n-1}}$ is a $(2^n+1)$-qubit state, the inequality \eqref{n24} leads us to
\begin{equation*}
[E_{q, s}(\Gamma_{A|B_0B_1\ldots B_{2^n-1}})]^\alpha
\geq\sum\nolimits_{j=0}^{2^n-1}\Big(\frac{(1+k^{\delta})^\alpha-1}{k^{\delta\alpha}}\Big)^{\omega_H(\vec{j})}[E_{q, s}(\Gamma_{A|B_j})]^\alpha,
\end{equation*}
where $\Gamma_{AB_j}$ is the two-qubit reduced state of $\Gamma_{AB_0B_1\ldots B_{2^n-1}}$ for $j=0,1,\ldots,2^n-1$. Thus, we have
\begin{align*}
[E_{q, s}(\rho_{A|B_0B_1\ldots B_{N-1}})]^\alpha =&[E_{q, s}(\Gamma_{A|B_0B_1\ldots B_{2^n-1}})]^\alpha \\[1mm]
\geq &\sum\nolimits_{j=0}^{2^n-1}\Big(\frac{(1+k^{\delta})^\alpha-1}{k^{\delta\alpha}}\Big)^{\omega_H(\vec{j})}[E_{q, s}(\Gamma_{A|B_j})]^\alpha \\[1mm]
=&\sum\nolimits_{j=0}^{N-1}\Big(\frac{(1+k^{\delta})^{\alpha}-1}{k^{\delta\alpha}}\Big)^{\omega_H(\vec{j})}[E_{q, s}(\rho_{A|B_j})]^\alpha.
\end{align*}
Since the separability $\Gamma_{A|B_0B_1\ldots B_{2^n-1}}$ with respect to the bipartition $AB_0\ldots B_{N-1}$ and $B_N\ldots B_{2^n-1}$ assures that $E_{q, s}\left(\Gamma_{A|B_0 B_1 \cdots B_{2^n-1}}\right)=E_{q, s}\left(\rho_{A|B_0 B_1 \cdots B_{N-1}}\right)$,
$E_{q, s}\left(\Gamma_{A|B_j}\right)=0$ for $j=N, \cdots , 2^n-1$,
and $\Gamma_{AB_j}=\rho_{AB_j}$ for each $j=0, \cdots , N-1$, we complete the proof.
\qed

Since $\Big(\frac{(1+k^\delta)^\alpha-1}{k^{\delta\alpha}}\Big)^{\omega_H(\vec{j})}\geq \alpha^{\omega_H(\vec{j})}$ for $\alpha \geq1$, we have the following relation,
\begin{equation*}
[E_{q, s}(\rho_{A|B_0B_1\ldots B_{N-1}})]^\alpha
\geq\sum\nolimits_{j=0}^{N-1}\Big(\frac{(1+k^\delta)^\alpha-1}
{k^{\delta\alpha}}\Big)^{\omega_H(\vec{j})}[E_{q, s}(\rho_{A|B_j})]^\alpha
\geq\sum\nolimits_{j=0}^{N-1}\alpha^{\omega_H(\vec{j})}[E_{q, s}(\rho_{A|B_j})]^\alpha
\end{equation*}
for any multiqubit state $\rho_{AB_0 B_1 \cdots B_{N-1}}$ and $\alpha \geq1$.
Therefore, our result in Theorem 1 is better than the inequality \eqref{n12}.

\textbf{Example 1}
Let us consider the generalized three-qubit state $|\psi\rangle_{ABC}$ in Schmidt decomposition \cite{c31,c32},
\begin{equation} \label{n33}
|\psi\rangle_{ABC}=\lambda_0|000\rangle+\lambda_1e^{i{\varphi}}|100\rangle+\lambda_2|101\rangle
+\lambda_3|110\rangle+\lambda_4|111\rangle,
\end{equation}
where $\lambda_i\geq0,~i=0, 1, 2, 3, 4$ and $\sum_{i=0}^4\lambda_i^2=1.$ From the definition of UE in \eqref{n1}, when $q=2$ and $s$ tends to 1, we have
\begin{align*}
&T_2(|\psi\rangle_{A|BC})=1-\T(\rho^2_A),\\
&T_2(\rho_{AB})=\min\sum_ip_iT_2(|\psi_i\rangle_{AB}),
\end{align*}
where $\rho_A=\T_{BC}(|\psi\rangle_{ABC}\langle\psi|)$, $\rho_{AB}=\T_C(|\psi\rangle_{ABC}\langle\psi|)$ and the minimum is taken over all possible pure state decompositions of
$\rho_{AB}=\sum_{i}p_i (|\psi_i \rangle_{AB} \langle \psi _i|)$ with $p_i\geq0$ and $\sum_{i}p_i=1$.  Then we get
$T_2(|\psi\rangle_{A|BC})=2\lambda_0^2(\lambda_2^2+\lambda_3^2+\lambda_4^2)$,
$T_2(\rho_{AB})=\lambda_0^2\lambda_2^2$ and $T_2(\rho_{AC})=2\lambda_0^2\lambda_3^2$.
Taking $\lambda_0=\frac{\sqrt{3}}{3}$, $\lambda_2=\frac{\sqrt{2}}{2}$, $\lambda_3=\frac{\sqrt{6}}{6}$, $\lambda_1=\lambda_4=0$ and choosing $k=\frac{3}{4}$ and $\delta=1$, we have
\begin{equation*}
 T^\alpha_2(|\psi\rangle_{A|BC})=(\frac{4}{9})^\alpha,
\end{equation*}
and the lower bound of \eqref{n18},
\begin{equation*}
y_1\equiv T^\alpha_2(\rho_{AB})+\Big(\frac{(1+k^\delta)^\alpha-1}{k^{\delta\alpha}}\Big)
T^\alpha_2(\rho_{AC})=(\frac{1}{6})^\alpha+(\frac{7}{27})^\alpha-(\frac{4}{27})^\alpha.
\end{equation*}
While the lower bound of \eqref{n12} is given by
\begin{equation*}
y_2\equiv T^\alpha_2(\rho_{AB})+\alpha T^\alpha_2(\rho_{AC})=(\frac{1}{6})^\alpha+\alpha(\frac{1}{9})^\alpha.
\end{equation*}
One verifies that our lower bound is tighter than the one in \cite{b26}, see Fig. 1.
\begin{figure}
  \centering
  \includegraphics[width=8cm]{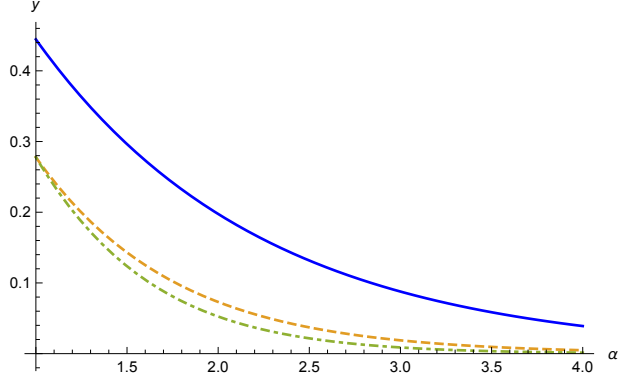}\\
  \caption{The solid line represents the UE of $|\psi\rangle_{ABC}$ as a function of $\alpha$. The dashed line is the lower bound $y_1$ from Theorem 1. The dot-dashed line represents the lower bound $y_2$ give in \cite{b26}.}
\end{figure}

The inequality \eqref{n18} of Theorem 1 can be further improved to be much tighter under certain conditions.

\begin{theorem}
For $\alpha\geq1$, $q\geq2$, $0\leq s \leq1$ and $qs\leq3$, as well as $0< k \leq1$ and $\delta\geq1$, we have for any multiqubit state $\rho_{AB_0\ldots B_{N-1}}$,
\begin{equation*}\label{n31}
[E_{q, s}(\rho_{A|B_0\ldots B_{N-1}})]^\alpha
\geq\sum\nolimits_{j=0}^{N-1}\Big(\frac{(1+k^\delta)^\alpha-1}{k^{\delta\alpha}}\Big)^j[E_{q, s}(\rho_{A|B_j})]^\alpha,
\end{equation*}
if $k^\delta E_{q, s}(\rho_{A|B_i})\geq\sum\nolimits_{j=i+1}^{N-1}E_{q, s}(\rho_{A|B_j})$
for $i=0,1,\ldots,N-2$.
\end{theorem}

{\sf [Proof]}
From the inequality \eqref{n9}, we only need to prove that
\begin{equation}
\left(\sum_{j=0}^{N-1}E_{q, s}\left(\rho_{A|B_j}\right)\right)^{\alpha}
\geq \sum_{j=0}^{N-1} \left( \frac{(1+k^\delta)^\alpha-1}{k^{\delta\alpha}}\right)^{j}\left(E_{q, s}\left(\rho_{A|B_j}\right)\right)^{\alpha}.\label{n26}
\end{equation}
We use mathematical induction on $N$. Note that the inequality \eqref{n10} in the proof of Theorem 1 assures that the inequality \eqref{n26} is true for $N = 2$.
Assume that the inequality \eqref{n26} holds for any positive integer less than $N$.

Since for any multiqubit state $\rho_{AB_0\ldots B_{N-1}}$,
\begin{equation*}
\left(\sum_{j=0}^{N-1}E_{q, s}\left(\rho_{A|B_j}\right)\right)^{\alpha}
=\left(E_{q, s}\left(\rho_{A|B_0}\right)\right)^{\alpha}
\left(1+\frac{\sum_{j=1}^{N-1}E_{q, s}\left(\rho_{A|B_j}\right)}
{E_{q, s}\left(\rho_{A|B_0}\right)} \right)^{\alpha}
\end{equation*}
and
\begin{equation*}
\left(1+\frac{\sum_{j=1}^{N-1}E_{q, s}\left(\rho_{A|B_j}\right)}
{E_{q, s}\left(\rho_{A|B_0}\right)} \right)^{\alpha}
\geq 1 + \left( \frac{(1+k^\delta)^\alpha-1}{k^{\delta\alpha}}\right) \left(\frac{\sum_{j=1}^{N-1}E_{q, s}\left(\rho_{A|B_j}\right)}
{E_{q, s}\left(\rho_{A|B_0}\right)}\right)^{\alpha},
\end{equation*}
we have
\begin{align*}
\left(\sum_{j=0}^{N-1}E_{q, s}\left(\rho_{A|B_j}\right) \right)^{\alpha}
\geq & \left(E_{q, s}\left(\rho_{A|B_0}\right)\right)^{\alpha}
+\left( \frac{(1+k^\delta)^\alpha-1}{k^{\delta\alpha}}\right) \left(\sum_{j=1}^{N-1}E_{q, s}\left(\rho_{A|B_j}\right) \right)^{\alpha} \nonumber\\
\geq & \sum_{j=0}^{N-1} \left( \frac{(1+k^\delta)^\alpha-1}{k^{\delta\alpha}}\right)^{j}\left(E_{q, s}\left(\rho_{A|B_j}\right)\right)^{\alpha},
\end{align*}
where the second inequality is based on the inductive hypothesis, which proves the theorem.
\qed

In fact, for any $\alpha \geq 1$, according to \eqref{n11} one has
\begin{align*}
[E_{q, s}(\rho_{A|B_0\ldots B_{N-1}})]^\alpha
\geq & \sum_{j=0}^{N-1} \left( \frac{(1+k^\delta)^\alpha-1}{k^{\delta\alpha}}\right)^{j}\left(E_{q, s}\left(\rho_{A|B_j}\right)\right)^{\alpha} \nonumber\\
\geq & \sum\nolimits_{j=0}^{N-1}\Big(\frac{(1+k^\delta)^\alpha-1}
{k^{\delta\alpha}}\Big)^{\omega_H(\vec{j})}[E_{q, s}(\rho_{A|B_j})]^\alpha.
\end{align*}
We can also obtain an upper bound for $[E_{q, s}(\rho_{A|B_0B_1\ldots B_{N-1}})]^\alpha$ when $\alpha<0$.

\begin{theorem}
For any multiqubit state $\rho_{AB_0\ldots B_{N-1}}$ with $E_{q, s}(\rho_{AB_i})\neq0$,
$i=0,1,\ldots,N-1$,
we have	
\begin{equation}\label{SC28}
[E_{q, s}(\rho_{A|B_0B_1\ldots B_{N-1}})]^\alpha
<\frac{1}{N}\sum\nolimits_{j=0}^{N-1}[E_{q, s}(\rho_{A|B_j})]^\alpha,
\end{equation}
for all $\alpha<0$ and  $q\geq2$, $0\leq s \leq1$ and $qs\leq3$.
\end{theorem}

{\sf [Proof]} Similar to the proof of Theorem 3 in \cite{jin}, for any three-qubit state we have
\begin{align}\label{SCREN1}
[E_{q, s}(\rho_{A|B_0B_1})]^\alpha
\leq & [E^{2}_{q, s}(\rho_{A|B_0})+ E^{2}_{q, s}(\rho_{A|B_1})]^{\frac{\alpha}{2}}\nonumber\\
=& [E_{q, s}(\rho_{A|B_0})]^\alpha \Big(1+\frac{E^{2}_{q, s}(\rho_{A|B_1})}{E^{2}_{q, s}(\rho_{A|B_0})}\Big)^{\frac{\alpha}{2}}\nonumber\\
<& [E_{q, s}(\rho_{A|B_0})]^\alpha,
\end{align}
where the first inequality is due to the fact that $\alpha<0$, and the second inequality is due to
$\Big(1+\frac{E^{2}_{q, s}(\rho_{A|B_1})}{E^{2}_{q, s}(\rho_{A|B_0})}\Big)^\alpha<1$.
From \eqref{SCREN1} and $[E_{q, s}(\rho_{A|B_0B_1})]^\alpha<[E_{q, s}(\rho_{A|B_1})]^\alpha$
we have
\begin{equation*}
[E_{q, s}(\rho_{A|B_0B_1})]^\alpha
<\frac{1}{2}\{[E_{q, s}(\rho_{A|B_0})]^\alpha+[E_{q, s}(\rho_{A|B_1})]^\alpha\}.
\end{equation*}
Thus, we obtain
\begin{align}\label{S23}
&[E_{q, s}(\rho_{A|B_0B_1\ldots B_{N-1}})]^\alpha \nonumber\\
< & \frac{1}{2}\Bigg\{\Big[E_{q, s}(\rho_{A|B_0})\Big]^\alpha+\Big[E_{q, s}(\rho_{A|B_1\ldots B_{N-1}})\Big]^\alpha\Bigg\}\nonumber\\
<& \frac{1}{2}\Big[E_{q, s}(\rho_{A|B_0})\Big]^\alpha+\Big(\frac{1}{2}\Big)^2\Big[E_{q, s}(\rho_{A|B_1})\Big]^\alpha
+\Big(\frac{1}{2}\Big)^2\Big[E_{q, s}(\rho_{A|B_2\ldots B_{N-1}})\Big]^\alpha \nonumber\\
<& \ldots \nonumber\\
<& \frac{1}{2}\Big[E_{q, s}(\rho_{A|B_0})\Big]^\alpha+\Big(\frac{1}{2}\Big)^2\Big[E_{q, s}(\rho_{A|B_1})\Big]^\alpha+\ldots
+\Big(\frac{1}{2}\Big)^{N-1}\Big[E_{q, s}(\rho_{A|B_{N-2}})\Big]^\alpha
+\Big(\frac{1}{2}\Big)^{N-1}\Big[E_{q, s}(\rho_{A|B_{N-1}})\Big]^\alpha.
\end{align}
Similarly, we can obtain a set of inequalities by the cyclic permutation of the indices $B_0$, $B_1,$ $\ldots$, $B_{N-1}$ in \eqref {S23}. Adding up these inequalities, we get
\eqref{SC28}.
\qed

\section{Tighter polygamy constraints of multiqubit entanglement in terms of UEoA}

In this section we consider the UEoA defined in \eqref{n2}. Dual to inequality \eqref{n18} in Theorem 1, we present a class of polygamy inequalities satisfied by the multiqubit UEoA.

\begin{theorem}
For any multiqubit state $\rho_{AB_0\ldots B_{N-1}}$, $0\leq \beta \leq1$, $1 \leq q \leq 2$, $-q^2+4q-3 \leq s\leq1$, and real numbers $0 < k \leq 1$, $\delta\geq1$
we have
\begin{equation}\label{n38}
[E^{a}_{q, s}(\rho_{A|B_0B_1\ldots B_{N-1}})]^\beta
\leq\sum\limits_{j=0}^{N-1}\Big(\frac{(1+k^\delta)^\beta-1}{k^{\delta\beta}}\Big)^{\omega_H(\vec{j})}[E^{a}_{q, s}(\rho_{A|B_j})]^\beta.
\end{equation}
\end{theorem}

{\sf [Proof] }
Without loss of generality, we assume that
\begin{equation}\label{n40}
k^\delta E^{a}_{q, s}(\rho_{A|B_j})\geq E^{a}_{q, s}(\rho_{A|B_{j+1}})\geq 0
\end{equation}
under suitable ordering of the subsystems $B_0, \ldots, B_{N-1}$,
where $j=0,1,\ldots,N-2$ and $0<k\leq1$, $\delta \geq 1$.
Due to the monotonicity of the function $f(x)=x^\beta$ for $0\leq \beta\leq1$ and the inequality \eqref{n8}, we only need to prove that
\begin{equation}\label{n39}
\left[\sum\limits_{j=0}^{N-1} E^{a}_{q, s}(\rho_{A|B_j})\right]^\beta
\leq\sum\limits_{j=0}^{N-1}\Big(\frac{(1+k^\delta)^\beta-1}
{k^{\delta\beta}}\Big)^{\omega_{H}(\vec{j})}
[E^{a}_{\alpha}(\rho_{A|B_j})]^\beta.
\end{equation}

Similar to the proof of Theorem 1, we first prove the inequality \eqref{n39} for the
case that $N = 2^n$ by using mathematical induction on $n$, and generalize the result to arbitrary $N$. For $n=1$ and a three-qubit state $\rho_{AB_0 B_1}$ with two-qubit reduced states $\rho_{AB_0}$ and $\rho_{AB_1}$, we have
\begin{align*}
[E^{a}_{q, s}(\rho_{A|B_0})+E^{a}_{q, s}(\rho_{A|B_1})]^\beta
=& [E^{a}_{q, s}(\rho_{A|B_0})]^\beta \Big(1+\frac{E^{a}_{q, s}(\rho_{A|B_1})}{E^{a}_{q, s}(\rho_{A|B_0})}\Big)^\beta \nonumber\\
\leq & [E^{a}_{q, s}(\rho_{A|B_0})]^\beta \left[1+\displaystyle\frac{(1+k^\delta)^\beta-1}{k^{\delta\beta}}\Bigg(\frac{E^{a}_{q, s}(\rho_{A|B_1})}{E^{a}_{q, s}(\rho_{A|B_0})}\Bigg)^\beta \right] \nonumber\\
=& [E^{a}_{q, s}(\rho_{A|B_0})]^\beta+\displaystyle\frac{(1+k^\delta)^\beta-1}{k^{\delta\beta}}[E^{a}_{q, s}(\rho_{A|B_1})]^\beta,
\end{align*}
where the inequality is due to the inequality \cite {LYY},
 \begin{equation*}
 	(1+x)^{\beta}\leq 1+\frac{(1+k^{\delta})^{\beta}-1}{k^{\delta\beta}}x^{\beta}, \label{n17}
 \end{equation*}
 where $k$ and $\delta$ are any real numbers satisfying $0<k \leq 1$ and $\delta\geq 1$, $0\leq x \leq k^{\delta}$ and $0\leq \beta \leq1$.

Next, we assume that the inequality \eqref{n39} is true for $N=2^{n-1}$ with $n\geq 2$, and consider the case of $N=2^n$.
For an $(N + 1)$-qubit state $\rho_{AB_0B_1\ldots B_{N-1}}$ with its two-qubit reduced states $\rho_{AB_j}$, $j=0,1,\ldots,N-1$, we have $E^{a}_{q, s}(\rho_{A|B_{j+2^{n-1}}})\leq k^{\delta2^{n-1}}E^{a}_{q, s}(\rho_{A|B_j})$ due to the inequality \eqref{n40}.
Then, we get
\begin{equation*}
0\leq\frac{\sum\nolimits_{j=2^{n-1}}^{2^n-1}E^{a}_{q, s}(\rho_{A|B_j})}{\sum\nolimits_{j=0}^{2^{n-1}-1}
E^{a}_{q, s}(\rho_{A|B_j})}\leq k^{\delta2^{n-1}}\leq k^\delta \leq 1
\end{equation*}
and
\begin{equation*}
\Bigg(\sum\nolimits_{j=0}^{N-1}E^{a}_{q, s}(\rho_{A|B_j})\Bigg)^\beta
=\Bigg(\sum\nolimits_{j=0}^{2^{n-1}-1}E^{a}_{q, s}(\rho_{A|B_j})\Bigg)^\beta
\Bigg(1+\frac{\sum_{j=2^{n-1}}^{2^n-1}E^{a}_{q, s}(\rho_{A|B_j})}{\sum_{j=0}^{2^{n-1}-1}E^{a}_{q, s}
(\rho_{A|B_j})}\Bigg)^\beta.
\end{equation*}
Hence,
\begin{equation}\label{n58}
\Bigg(\sum\nolimits_{j=0}^{N-1}E^{a}_{q, s}(\rho_{A|B_j})\Bigg)^\beta
\leq \Bigg(\sum\nolimits_{j=0}^{2^{n-1}-1}E^{a}_{q, s}(\rho_{A|B_j})\Bigg)^\beta
+\displaystyle\frac{(1+k^\delta)^\beta-1}{k^{\delta\beta}}\Bigg(\sum\nolimits_{j=2^{n-1}}^{2^n-1}E^{a}_{q, s}(\rho_{A|B_j})\Bigg)^\beta.
\end{equation}
Because each summation on the right-side of the inequality \eqref{n58} is a summation of $2^{n-1}$ terms,
the induction hypothesis assures that
$$
\Bigg(\sum\nolimits_{j=0}^{2^{n-1}-1}E^{a}_{q, s}(\rho_{A|B_j})\Bigg)^\beta
\leq \sum\nolimits_{j=0}^{2^{n-1}-1}\Big(\frac{(1+k^\delta)^\beta-1}
{k^{\delta\beta}}\Big)^{\omega_H(\vec{j})}
[E^{a}_{q, s}(\rho_{A|B_j})]^\beta
$$
and
$$
\Bigg(\sum\nolimits_{j=2^{n-1}}^{2^n-1}E^{a}_{q, s}(\rho_{A|B_j})\Bigg)^\beta
\leq \sum\nolimits_{j=2^{n-1}}^{2^n-1}\Big(\frac{(1+k^\delta)^\beta-1}
{k^{\delta\beta}}\Big)^{\omega_H(\vec{j})-1}[E^{a}_{q, s}(\rho_{A|B_j})]^\beta.
$$
Therefore,
$$
\Bigg(\sum\nolimits_{j=0}^{2^n-1}E^{a}_{q, s}(\rho_{A|B_j})\Bigg)^\beta
\leq \sum\nolimits_{j=0}^{2^n-1}\Big(\frac{(1+k^\delta)^\beta-1}{k^{\delta\beta}}\Big)^{\omega_H(\vec{j})}[E^{a}_{a, s}(\rho_{A|B_j})]^\beta.
$$
Concerning the $(2^n+1)$-qubit state \eqref{n25}, we have
\begin{align*}
[E^{a}_{q, s}(\rho_{A|B_0B_1\ldots B_{N-1}})]^\beta =&[E^{a}_{q, s}(\Gamma_{A|B_0B_1\ldots B_{2^n-1}})]^\beta \nonumber\\
\leq & \sum\nolimits_{j=0}^{2^n-1}\Big(\frac{(1+k^\delta)^\beta-1}{k^{\delta\beta}}\Big)^{\omega_H(\vec{j})}[E^{a}_{q, s}(\Gamma_{A|B_j})]^\beta \nonumber\\
=& \sum\nolimits_{j=0}^{N-1}\Big(\frac{(1+k^\delta)^\beta-1}{k^{\delta\beta}}\Big)^{\omega_H(\vec{j})}[E^{a}_{q, s}(\rho_{A|B_j})]^\beta,
\end{align*}
which proves the theorem.
\qed

Since $\frac{(1+k^\delta)^\beta-1}{k^{\delta\beta}}\leq\beta$ for $0\leq \beta \leq1$, it is easy to see that \eqref{n38} is tighter than \eqref{n13}.

\textbf{Example 2}
We consider again the three-qubit state $|\psi\rangle_{ABC}$ in \eqref{n33}.
From the definition \eqref{n2} of UEoA, when $q=2$ and $s$ tends to 1, we have
\begin{align*}
&T^a_2(|\psi\rangle_{A|BC})=1-\T(\rho^2_A),\\
&T^a_2(\rho_{AB})=\max\sum_ip_iT_2(|\psi_i\rangle_{AB}),
\end{align*}
where the maximum is taken over all possible pure state decompositions of
$\rho_{AB}=\sum_{i}p_i (|\psi_i \rangle_{AB} \langle \psi _i|)$ with $\sum_{i}p_i=1$ and $\rho_A=\T_{BC}(|\psi\rangle_{ABC}\langle\psi|)$ and $\rho_{AB}=\T_C(|\psi\rangle_{ABC}\langle\psi|)$. Then we get
$T^a_2(|\psi\rangle_{A|BC})=2\lambda_0^2(\lambda_2^2+\lambda_3^2+\lambda_4^2)$,
$T^a_2(\rho_{AB})=2\lambda_0^2(\lambda_2^2+\lambda_4^2)$ and
$T^a_2(\rho_{AC})=2\lambda_0^2(\lambda_3^2+\lambda_4^2)$.
Taking $\lambda_0=\frac{\sqrt{3}}{3}$, $\lambda_2=\frac{\sqrt{2}}{2}$, $\lambda_3=\frac{\sqrt{6}}{6}$, $\lambda_1=\lambda_4=0$, for the case of $k=\frac{2}{3}$ and  $\delta=1$ we get
\begin{equation*}
 [T^a_2(|\psi\rangle_{A|BC})]^\beta=(\frac{4}{9})^\beta,
\end{equation*}
and the upper bound of \eqref{n40},
\begin{equation*}
y_3\equiv [T^a_2(\rho_{AB})]^\beta+\Big(\frac{(1+k^\delta)^\beta-1}{k^{\delta\beta}}\Big)
[T^a_2(\rho_{AC})]^\beta=(\frac{1}{3})^\beta+(\frac{5}{18})^\beta-(\frac{1}{6})^\beta.
\end{equation*}
While the upper bound of \eqref{n13} is given by
\begin{equation*}
y_4\equiv [T^a_2(\rho_{AB})]^\beta+\beta [T^a_2(\rho_{AC})]^\beta=(\frac{1}{3})^\beta+\beta(\frac{1}{9})^\beta.
\end{equation*}
Fig. 2 shows that our result is better than \eqref {n13} from \cite{b26}.
\begin{figure}
  \centering
  \includegraphics[width=8cm]{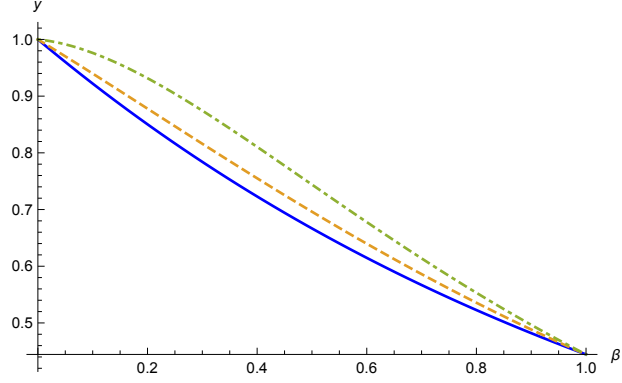}\\
  \caption{The solid line is the UEoA of $|\psi\rangle_{ABC}$ with respect to $\beta$. The dashed line is the upper bound $y_3$ from our theorem. The dot-dashed line is the upper bound $y_4$ give in \cite{b26}.}
\end{figure}

The inequality \eqref{n38} of Theorem 4 can be again improved to a class of tighter polygamy inequalities under certain conditions on the two-qubit entanglement of assistance. The proof is similar to the one of Theorem 2.

\begin{theorem}
For $0 \leq \beta \leq 1$, $1 \leq q \leq 2$ and $-q^2+4q-3 \leq s \leq1$, as well as $0< k \leq1$ and $\delta \geq 1$, we have for any multiqubit state $\rho_{AB_0\ldots B_{N-1}}$,
\begin{equation} \label{ps}
[E^{a}_{q, s}(\rho_{A|B_0\ldots B_{N-1}})]^\beta
\leq \sum\nolimits_{j=0}^{N-1}\Big(\frac{(1+k^\delta)^\beta-1}{k^{\delta\beta}}\Big)^j[E^{a}_{q, s}(\rho_{A|B_j})]^\beta,
\end{equation}
if
\begin{equation} \label{ps1}
k^{\delta}E^{a}_{q, s}(\rho_{A|B_i})\geq\sum\nolimits_{j=i+1}^{N-1}E^{a}_{q, s}(\rho_{A|B_j})
\end{equation}
for $i=0,1,\ldots,N-2$.
\end{theorem}

Since $\omega_H(\vec{j})\leq j$, for $0\leq\beta\leq1$ we have
\begin{align*}
[E^{a}_{q, s}(\rho_{A|B_0\ldots B_{N-1}})]^\beta
\leq & \sum_{j=0}^{N-1} \left( \frac{(1+k^\delta)^\beta-1}{k^{\delta\beta}}\right)^{j}\left(E^{a}_{q, s}\left(\rho_{A|B_j}\right)\right)^{\beta} \nonumber\\
\leq & \sum\nolimits_{j=0}^{N-1}\Big(\frac{(1+k^\delta)^\beta-1}{k^{\delta\beta}}\Big)^{\omega_H(\vec{j})}[E^{a}_{q, s}(\rho_{A|B_j})]^\beta.
\end{align*}
Thus, the inequality \eqref{ps} of Theorem 5 is tighter than the inequality \eqref{n38} of Theorem 4 for $0\leq\beta\leq1$ and any multiqubit state  $\rho_{AB_0\ldots B_{N-1}}$ satisfying the condition \eqref{ps1}.

\section{conclusion}
Quantum entanglement plays a crucial role in quantum information processing such as quantum communication. The entanglement monogamy and polygamy relations are two fundamental properties of multipartite entangled states. In this article, by using the Hamming weight of binary vector related to the distribution of subsystems, we have provided a class of monogamy inequalities based on the $\alpha$th ($\alpha\geq1$) power of UE. We have also provided a class of polygamy inequalities in terms of the $\beta$-th ($0\leq \beta \leq 1$) power of UEoA.
Applying these results to specific quantum correlations, such as entanglement of formation, R\'enyi-$q$ entanglement and Tsallis-$q$ entanglement, we can obtain the corresponding monogamy and polygamy relations. Furthermore, our classes of monogamy and polygamy inequalities are tighter than the existing ones, which give rise to finer characterizations of the entanglement shareability and distribution among the multiqubit subsystems. Our results may help to understand further the monogamy and polygamy nature of multiparty quantum entanglement.

\bigskip
\noindent{\bf Acknowledgments}\, \, This work is supported by NSFC (Grant No. 12075159), Beijing Natural Science Foundation (Z190005), Academy for Multidisciplinary Studies, Capital Normal University, the Academician Innovation Platform of Hainan Province, and Shenzhen Institute for Quantum Science and Engineering, Southern University of Science and Technology (No. SIQSE202001).

\end{document}